\documentclass[preprint,aps]{revtex4}

\usepackage{graphicx}
\usepackage{dcolumn}
\usepackage{bm}

\begin{document}

\title{Probe Noncommutative Space-Time Scale Using $\gamma\gamma \to Z$ At ILC}

\author{Xiao-Gang He$^{1,2}$, Xue-Qian Li$^2$}
\affiliation{$^{1}$NCTS/TPE, Department of Physics, National
Taiwan
University, Taipei\\
$^2$Department of Physics, Nankai University, Tianjin}

\date{\today} 

\begin{abstract}
In the standard model  production of on-shell Z boson at a photon
collider (or Z decays into $\gamma\gamma$) is strictly forbidden
by angular momentum conservation and Bose statistics (the Yang's
Theorem). In the standard model with noncommutative space-time
this process can occur. Therefore this process provides an
important probe for the noncommutative space-time. The
$\gamma\gamma$ collision  at the ILC by laser backscattering of
the electron and positron beams offers an ideal place to carry out
such a study. Assuming an integrated luminosity of 500 fb$^{-1}$,
we show that the constraint which can be achieved on $\Gamma(Z\to
\gamma \gamma)$ is three to four orders of magnitude better than
the current bound of $5.2\times 10^{-5}$ GeV. The noncommutative
scale can be probed up to a few TeVs.
\end{abstract}


\maketitle

\newpage

The property of space-time has fundamental importance in
understanding the laws of nature. Noncommutative quantum field
theory, which modifies the space-time commutation relations
provides an alternative to the ordinary quantum field theory, may
shed some light on the detailed structure of space-time. The idea
that the space-time coordinates may not commute has a long
history\cite{hist1,syn}. In recent years, noncommutative
space-time has found a natural origin in string
theories\cite{sw,ho,rev}. A simple and commonly studied
noncommutative quantum field theory is based on the following
commutation relation of space-time,
\begin{eqnarray}
[\hat x_\mu, \hat x_\nu] = i \Theta_{\mu\nu}, \label{commu}
\end{eqnarray}
where $\hat x_\mu$ is the noncommutative space-time coordinates.
$\Theta_{\mu\nu}$ is a constant, real, anti-symmetric matrix, and
has $mass^{-2}$ dimension. The size of the
$1/\sqrt{|\Theta_{\mu\nu}|}$ represents the noncommutative scale
$\Lambda_{NC}$.

In the literature there are extensive studies on the mathematic
structure of noncommutative space-time, but less studies on
measurable physical consequences of the noncommutative models for
strong and electroweak processes\cite{rev}. To know if
noncommutative space-time has anything to do with nature, emphasis
should be placed more on physical consequences. Although there are
several studies for the constraints on noncommutative scale
$\Lambda_{NC}$\cite{constraint1,cons2,calmet}, most of the
analyses were based on $U(1)$ noncommutative gauge theory which is
not yet a consistent noncommutative model of the strong and
electroweak interactions. At present the constraints are fairly
loose with $\Lambda_{NC}$ only being limited to larger than a few
TeV\cite{calmet}. The noncommutative scale may be accessible at
near future colliders, such as the Large Hadron Collider (LHC) and
the International Linear Collider (ILC). In this work based on
consistent formulation of noncommutative standard model (NCSM), we
study the process $\gamma\gamma \to Z$ at ILC and new constraint
on the noncommutative scale may be achieved.

In the SM, $\gamma\gamma \to Z$ is strictly forbidden, if all
particles are on-shell, by angular momentum conservation and Bose
statistics (the Yang's Theorem)\cite{yang}, but can occur in
noncommutative version of the standard model (NCSM).  Note that if
one of the photon is off-shell, there can be non-zero self
coupling of the type $Z\gamma \gamma^*$\cite{hagi} even without
noncommutative space-time, and can be studied in process like
$p\bar p \to \gamma^* \to \gamma Z \to \gamma l \bar l$\cite{d0}.
Therefore production of on-shell $Z$ boson at $\gamma\gamma$
collider can provide an interesting test for the NCSM and
therefore the nature of space-time. We will show that the
constraint can be achieved on $\Gamma(Z\to \gamma\gamma)$ is
several orders of magnitude better than the present bound of
$5.2\times 10^{-5}$ GeV at 95\% C.L.\cite{pdg}. $\Lambda_{NC}$ can
be probed to a few TeV which is much better than that can be
extracted from present bound on $\Gamma(Z\to
\gamma\gamma)$\cite{wess}.

Quantum field theory based on the commutation relation in
eq.(\ref{commu}) can be easily studied using the Weyl-Moyal
correspondence, i.e.,  replacing the product of two fields $A(\hat
x)$ and $B(\hat x)$ with NC coordinates by the star ``*'' product:
\begin{eqnarray}
A(\hat x) B(\hat x) \to \hat A(x)*\hat B(x) = Exp[i{1\over 2}
\Theta_{\mu\nu}
\partial_{x}^{\mu}\partial_{y}^{\nu}]A(x)B(y)|_{x=y}.
\end{eqnarray}
Here fields with and without `hat' indicate the fields in
noncommutative space-time and ordinary space-time, respectively.

Promotion of the usual space-time coordinates $x_\mu$ to the
noncommutative space-time coordinates $\hat x_\mu$ has very
interesting consequences\cite{my}. We denote the noncommutative
gauge field to be $\hat A_\mu = \hat A^a_\mu T^a$ of a group with
generators normalized as $Tr(T^aT^b)= \delta^{ab}/2$. In
noncommutative space-time two consecutive local gauge
transformations $\hat \alpha$ and $\hat \beta$ of the type
$\delta_\alpha \hat \Psi = i \hat \alpha
* \hat \Psi$, with the  matter field $\hat \Psi$ transforming as a
fundamental representation of the gauge group, is given by $
(\delta_\alpha \delta_\beta - \delta_\beta \delta_\alpha) = (\hat
\alpha*\hat \beta - \hat \beta *\hat \alpha)$. This commutation
relation is consistent with the $U(N)$ Lie algebra, but not
consistent with the $SU(N)$ Lie algebra since it cannot be reduced
to the matrix commutators of the $SU(N)$ generators. Also note
that even with the $U(1)$ group the above consecutive
transformations do not commute implying that the charge for a
$U(1)$ gauge theory is limited to only three possible values which
can be normalized to 1, 0, -1.

The above properties pose difficulties in constructing
noncommutative standard model for the strong and electroweak
interactions because the standard gauge group contains $SU(3)_C$
and $SU(2)_L$ which cannot be naively gauged with noncommutative
space-time. Also the charges of $U(1)_Y$ are not just 1, 0, -1,
but some of them are fractionally charged after normalizing the
right-handed electron to have -1 hypercharge, such as, 1/6, 1/2,
2/3, -1/3 for left-handed quarks, left-handed leptons,
right-handed up and down quarks, respectively. This is the so
called charge quantization problem. However all these difficulties
can be overcome with the use of the Seiberg-Witten (SW)\cite{sw}
map which maps noncommutative gauge field  to ordinary commutative
gauge field. A consistent noncommutative $SU(N)$ gauge theory can
be constructed by expanding $\hat \alpha$ in powers of $\Theta$
with $\hat \alpha = \alpha + \alpha^{(1)}_{ab}:T^aT^b: + ... +
\alpha^{(n-1)}_{a_1...a_n}:T^{a_1}...T^{a_n}:...$ to form a closed
envelop algebra. Here the symbol `$:T^{a_1}...T^{a_n}:$' means
totally symmetric in exchanging $a_i$. Detailed description of the
method can be found in Ref.\cite{ncsm}. One can then expand gauge
and mater fields in powers of $\Theta$ to have a consistent
$SU(N)$ gauge theory order by order in $\Theta$. To the first
order in $\Theta$, one has\cite{ncsm}
\begin{eqnarray}
\hat A_\mu = A_\mu - {1\over 4}g_N\Theta^{\alpha\beta} \{A_\alpha,
\partial_\beta A_\mu + F_{\beta\mu}\}. \label{field}
\end{eqnarray}
The above gauge field would generate new terms in the interaction
Lagrangian compared with the ordinary $SU(N)$ gauge theory. For
example the term $-(1/2)Tr(F_{\mu\nu}F^{\mu\nu})$ in the
Lagrangian for an $SU(N)$ gauge field will become, to the first
order in $\Theta$\cite{ncsm},
\begin{eqnarray}
L &=& -{1\over 2} TrF_{\mu\nu}  F^{\mu\nu} + g_N \Theta^{\mu\nu}
{1\over 4} Tr[ F_{\mu\nu}F_{\rho\sigma}F^{\rho\sigma} - 4 F_{\mu
\rho}F_{\nu\sigma}F^{\rho\sigma}]. \label{nc}
\end{eqnarray}
The SW map can also cure the charge quantization problem by
associating a gauge field $\hat A_\mu^{(n)}$ for the a matter
field $\psi^{(n)}$ with $U(1)$ charge $gQ^{(n)}$, $\hat
A_\mu^{(n)} = A_\mu -(g Q^{(n)}/4)\Theta^{\alpha\beta} \{A_\alpha,
\partial_\beta A_\mu + F_{\beta\mu}\}$, where $A_\mu$ is
the gauge field of $U(1)$ in ordinary space-time.
 With help of the SW map
a specific method to construct NCSM and grand unified theories
have been developed\cite{ncsm,su5,su3,so10}.

The new terms in eq.(\ref{nc}) when applied to the NCSM, will
generate terms inducing $Z-\gamma-\gamma$ interaction. These terms
can be parameterized as
\begin{eqnarray}
L_{Z\gamma\gamma} = e g_{Z\gamma\gamma} \Theta^{\alpha\beta} (8
Z_{\mu \alpha} A_{\nu \beta}A^{\mu\nu} + 4 A_{\mu
\alpha}A_{\nu\beta} Z^{\mu\nu}- 2 A_{\alpha \beta} A_{\mu\nu}
Z^{\mu\nu} - Z_{\alpha\beta} A_{\mu\nu} A^{\mu\nu}).
\end{eqnarray}

In NCSM, $g_{Z\gamma\gamma}$ is not uniquely determined due to the
need of introducing a gauge field for each matter field with
different $U(1)_Y$ charge to solve the charge quantization
problem\cite{ncsm}. This is because that as summing over different
$U(1)_Y$ gauge fields for all matter fields to give the kinetic
energy, even though the first term in eq.(\ref{nc}) is fixed with
the right normalization, the triple gauge field terms are not
fixed. This problem may be solved by obtaining low energy NCSM
from grand unified theories such as noncommutative
SO(10)\cite{so10}, SU(5)\cite{su5} grand unification and
$SU(3)^3=SU(3)_C\times SU(3)_L\times SU(3)_R$
trinification\cite{su3} theories where there is no $U(1)$ charge
quantization problem to start with.  In noncommutative SO(10)
grand unification, due to the same reason this theory is anomaly
free, the triple gauge coupling is automatically zero. Therefore
in this model $\gamma\gamma\to Z$ cannot occur. Naively,
noncommutative $SU(5)$ grand unification can fix the triple gauge
boson couplings\cite{su5}. However, in this model, there are
several different multiplets for fermion and Higgs
representations, $\bar 5$, 10, 24 and etc., one needs to associate
different gauge fields with them which lead to a similar problem
of non-uniqueness of triple gauge boson couplings for different
$U(1)_Y$ gauge fields in the NCSM\cite{so10}. $SU(5)$ is not truly
a unified model in noncommutative space-time. Unique non-trivial
triplet gauge boson couplings can be generated in noncommutative
trinification model\cite{su3}. In this model, the fermion and
Higgs representations are all in the 27-representation of the
gauge group resulting in fixed triple gauge boson couplings. The
coupling $e g_{Z\gamma\gamma}$ in $SU(3)^3$ is given, at the
unification scale, by\cite{su3}
\begin{eqnarray}
e g_{Z\gamma\gamma} &=&-{g_U\over 16 \sqrt{15}} {4\over
5}\sin\theta_W (1 + {19\over 4}\cos 2\theta_W).
\end{eqnarray}
Using the normalization $g_Y = \sqrt{3/5}g_U$, and running down to
energy scale $\mu = m_Z$, we have $e g_{Z\gamma\gamma}=
-5.58\times 10^{-3}$. In rest of the discussions we will use
noncommutative $SU(3)^3$ as an illustration to show how the limit
on the noncommutative scale can be determined
 using $\gamma\gamma \to Z$ at the photon collision mode of ILC.

The matrix element for on-shell $\gamma(k)\gamma(k') \to Z(p)$ in
momentum space after symmetrizing the two photons is given by
\begin{eqnarray}
M = -i e g_{Z\gamma\gamma} 16 &[&k\cdot k'( k'\cdot \epsilon^*_Z
\epsilon\cdot \Theta\cdot \epsilon' + \epsilon'\cdot \epsilon^*_Z
k\cdot \Theta \epsilon' + \epsilon'\cdot \epsilon^*_Z k'\cdot
\Theta\cdot
\epsilon)\nonumber\\
& +& k\cdot \Theta \cdot k' ( k'\cdot \epsilon \epsilon'\cdot
\epsilon^*_Z - k\cdot \epsilon' \epsilon\cdot \epsilon^*_Z +
\epsilon\cdot \epsilon' k\cdot \epsilon^*_Z)],
\end{eqnarray}
where $a\cdot \Theta \cdot b = a_\alpha \Theta^{\alpha\beta}
b_\beta$.

With the above amplitude we obtain
\begin{eqnarray}
&&\sigma(\gamma\gamma \to Z, s) = 6 \pi^2 {m_Z \Gamma(Z\to
\gamma\gamma)\over m^2_Z}\delta(s - m^2_Z),\nonumber\\
&&\Gamma(Z\to \gamma\gamma) = {4\over 3}\alpha_{em}
g^2_{Z\gamma\gamma} m^5_Z (\Theta^2_S + {7\over 3} \Theta^2_T),
\end{eqnarray}
where $\Theta_T^2 = \theta^2_{01} + \theta^2_{02} + \theta^2_{03}$
and $\Theta^2_S = \theta^2_{12} + \theta^2_{13} + \theta^2_{23}$.
The expression for $\Gamma(Z\to \gamma\gamma)$ agrees with that
obtained in Ref.\cite{wess}.

We would like to emphasize the fact that in the usual SM the
process $\gamma\gamma \to Z$ is strictly forbidden makes it
possible to consistently test models which are based on
noncommutative space-time and expanded to the first order in
$\Theta$ in terms of the cross section of $\gamma\gamma \to Z$. It
is interesting to remark that if a process has SM contribution,
there can be two types of terms proportional to $\Theta^2$ for
cross section or decay rate, one from the square of the
interaction expanded to the first order in $\Theta$ and another
from interference of the usual SM and second order term in
$\Theta$ in NCSM. One then needs to expand the NCSM to second
order in $\Theta$ to have a consistent test. This can in principle
be carried out using the SW map described earlier but would
complicate the analysis. Tests of NCSM to the first order in
$\Theta$ are, however, still possible in this case if one uses
observable which is at the first order in $\Theta$, such as some
asymmetries in polarization of spin or certain angular
distribution of a particular particle involved\cite{Oh}. Here we
concentrate on the simple on-shell process $\gamma\gamma \to Z$.

Convoluting the energies of the two photon beams produced by using
the laser backscattering technique\cite{laser} on the electron and
positron beams in an $e^+ e^-$ collider with center of mass frame
energy $\sqrt{s}$, we have
\begin{eqnarray}
\sigma_c = \int^{x_{max}}_{x_{min}} d x_1 \int^{x_{max}}_{x_{min}}
dx_2 \sigma(\gamma\gamma \to Z, x_1x_2 s) F(x_1)F(x_2) =
I(m^2_Z/s) 6\pi^2{m_Z\Gamma(Z\to \gamma\gamma)\over m^4_Z},
\end{eqnarray}
where
\begin{eqnarray}
I(y) = \int^{x_{max}}_{y/x_{max}} dx {y\over x} F(x) F({y\over
x}),
\end{eqnarray}
with $y= m^2_Z/s$, and $x_{max} = \xi/(1+\xi)$ with $\xi =
2(1+\sqrt{2})$. The function $F(x)$ is given by
\begin{eqnarray}
&&F(x) = {1\over D(\xi)} ( 1-x +{1\over 1-x} - {4x\over \xi(1-x)} +
{4x^2\over \xi^2(1-x)^2}),\nonumber\\
&&D(\xi) = (1-{4\over \xi} - {8\over \xi^2})
\ln(1+\xi) + {1\over 2} + {8\over \xi}
- {1\over 2 (1+\xi)^2}.
\end{eqnarray}

Note that the function $I(y)$ is a function of $m^2_Z/s$ only. The
model-dependent part resides purely in the expression for
$\Gamma(Z\to \gamma\gamma)$. In Fig. 1 we show $I(y)$ as a
function of $y$. We see that for a large range of $m^2_Z/s$,
$I(y)$ is sizeable. The ILC of energy between 120 to 250 GeV can
be very useful for the purpose of studying $\gamma \gamma \to Z$.
When energy becomes higher the cross section goes down. If there
is a $Z'$ particle with a mass of a few hundred GeV, ILC of energy
around several hundred GeV to one TeV would be an excellent place
to look for $Z'$ via the process $\gamma \gamma \to Z'$.

\begin{figure}[htb]
\begin{center}
\includegraphics[width=8cm]{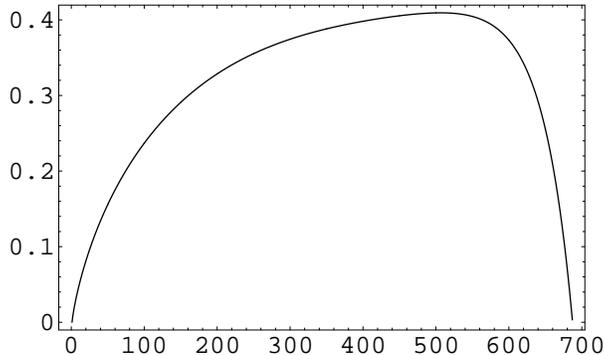}
\end{center}
\caption{ The function $I(y)$. The horizontal axis is $10^3y$.}
\end{figure}

The proposed ILC energy will be in the range from several hundred
GeV to TeV, and therefore can be an ideal place to study $\gamma
\gamma\to Z$. We list reachable upper bounds (using current bound
$5.2\times 10^{-5}$ GeV\cite{pdg} on the decay rate $\Gamma(Z\to
\gamma\gamma)$) on signal event number and the decay rate in Table
1, assuming an integrated luminosity of 500 fb$^{-1}$. We see that
if a theory gives $\Gamma$ for $Z\to \gamma\gamma$ close to the
current upper limit of $ 5.2 \times 10^{-5}$ GeV, one would see
more than $10^{5}$ events. If no events are seen, this information
would be translated into a bound on the rate $\Gamma$ of $Z\to
\gamma\gamma$ to be less than a few times of $ 10^{-10}$ GeV. This
is much better than the constraint obtained before\cite{wess}.
Even assuming an efficiency as low as 1\%, one can still set an
upper bound of $\Gamma < 10^{-8}$ GeV which is still more than
three orders of magnitude better than the current bound.

One can obtain the bound on the noncommutative scale
$\Lambda_{NC}$ from the bound on the event rate for $\gamma \gamma
\to Z$. We list the upper limits on the scales $\Lambda_S =
1/\sqrt{\Theta^2_S}$ and $\Lambda_T = 1/\sqrt{\Theta^2_T}$ in the
last two rows of Table 1. We see that the noncommutative scale can
be probed up to a few TeV. If the efficiency is lowered to 1\%,
the noncommutative scale can still be probed up to 1.5 TeV.

There are other three gauge boson interactions involving other
particles, such as $\gamma$, Z, and gluon g, which are not present
in the SM, but can exist in NCSM such as $Z-g-g$, $\gamma-g-g$,
$\gamma-\gamma-\gamma$, $Z-Z-\gamma$ and $Z-Z-Z$. Experimental
studies of other processes at LHC and ILC can also provide further
information about noncommutative space-time extension of the SM.
An interesting process similar to the process $\gamma\gamma \to Z$
discussed in this paper is gluon fusion into a Z-boson, i.e. $g g
\to Z$ at LHC by studying $pp\to Z X$. Note that this process is
not allowed in $SU(3)^3$\cite{su3} and nor in $SO(10)$\cite{so10}.
It is of course much more complicated to carry out such a study to
identify the noncommutative effects because $pp\to Z X$ can also
occur in the usual SM. More detailed studies are needed to isolate
the SM background and the NC effects.

\begin{table}[htb]
\begin{center}
\begin{tabular}{|c|c|c|c|c|c|} \hline
$\sqrt{s}$ (GeV)&120 & 200  &250 &500 &1000\\\hline
$I(m^2_Z/s)$&0.397&0.333&0.275&0.120&0.043\\\hline
 Upper limit on event number&$3.13\times 10^5$&$2.65\times 10^5$&$2.19\times
10^5$& $0.95\times 10^5$&$0.34\times 10^5$\\\hline Upper bound on
$\Gamma$/GeV &$1.66\times 10^{-10}$ &$1.96\times 10^{-10}$&
$2.38\times 10^{-10}$& $5.45\times 10^{-10}$&$1.51\times
10^{-9}$\\ \hline $SU(3)^3$: $\Lambda_S$ (TeV)&
4.72&4.53&4.31&3.51&2.72\\\hline $SU(3)^3$: $\Lambda_T$
(TeV)&5.83&5.59&5.33&4.33&3.36\\\hline
\end{tabular}
\end{center}
\caption{Upper limit on event number for $e^+e^-\to \gamma\gamma
\to Z$ (with current bound $5.2\times 10^{-5}$ GeV on $\Gamma(Z\to
\gamma\gamma$), and upper bound on $\Gamma(Z\to \gamma\gamma)$. In
obtaining these bounds the integrated luminosity is assumed to be
500 fb$^{-1}$. \label{fig-1}}
\end{table}

To summarize, we have studied the process $\gamma\gamma \to Z$
which is strictly forbidden in the standard model of strong and
electroweak interactions, but allowed in the noncommutative
space-time standard model. We have shown that the $\gamma\gamma$
collision mode at the ILC by laser backscattering of the electron
and positron beams with an integrated luminosity of 500 fb$^{-1}$
can obtain a constraint on $\Gamma(Z\to \gamma \gamma)$ three to
four orders of magnitude better than the current bound of
$5.2\times 10^{-5}$ GeV. The noncommutative scale can be probed up
to a few TeV. The process $\gamma\gamma\to Z$ at ILC can be a very
powerful test for NCSM.

\noindent {\bf Acknowledgments}: This work was supported in part
by NSC and NNSF.


\begin{references}
\bibitem{hist1} Letter of Heisenberg to Peierls (1930), Wolfgang
Pauli, Scientific Correspondence, Vol. II, p. 15, edited by Karl
von Meyenn (Springer-Verlag, 1985).

\bibitem{syn} H.S. Snyder, Phys. Rev. {\bf 71}, 38(1947).

\bibitem{sw} J. Seiberg and E. Witten, JHEP 9909, 032(1999).

\bibitem{ho}
C.~S.~Chu and P.~M.~Ho,
 Nucl.\ Phys.\ B {\bf 550}, 151 (1999)
 [arXiv:hep-th/9812219].
V.~Schomerus,
 JHEP {\bf 9906}, 030 (1999)
 [arXiv:hep-th/9903205].
C.~S.~Chu and P.~M.~Ho,
 Nucl.\ Phys.\ B {\bf 568}, 447 (2000)
 [arXiv:hep-th/9906192].

\bibitem{rev} M. Douglas and N.A. Nebrasov, Rev. Mod. Phys. {\bf
73}, 977(2001).

\bibitem{constraint1} J. Hewett, F. Petriello, T. Rizzo,
Phys. Rev. {\bf D 64}, 075012(2001); P. Mathew, Phys. Rev. {\bf D
63}, 075007(2001); S. Baek et al., Phys. Rev. {\bf D 64},
056001(2001), H. Grosse and Y. Liao, Phys. Rev. {\bf D 64},
115007(2001); S. Godgrey and M. Doncheski, Phys. Rev. {\bf D 65},
015005(2002).

\bibitem{cons2} S.M. Carrol et al., Phys. Rev. Lett. {\bf 87},
141601(2001); C.E. Carlson, C.D. Carone and R.F. Lebed, Phys.
Lett. {\bf 518}, 201(2001).

\bibitem{calmet} X. Calmet, Eur. J. Phys. {\bf C 41}, 269(2005).

\bibitem{yang} C.N. Yang, Phys. Rev. {\bf 77}, 242(1950).

\bibitem{hagi}K. Hagiwara et al., Nucl. Phys. {\bf B257},
531(1987); G. J. Gounaris, J. Layssac and F.M. Renard, Phys. Rev.
{\bf D62}, 073012(2000).

\bibitem{d0} V.M. Abazov et al., D0 Collaboration, Phys. Rev.
Lett. {\bf 95}, 051802(2005).

\bibitem{pdg} Particle Data Group, Phys. Lett. {\bf B592}, 1(2004).

\bibitem{wess} W. Behr et al., Eur. J. Phys. {\bf C 29},
441(2003).

\bibitem{my} M. Hayakawa, Phys. Lett. {\bf B 478}, 394(2000); K.
Matsubara, Phys. Lett. {\bf B 482}, 417(2000).



\bibitem{ncsm} J. Madore et al., Eur. J. Phys. {\bf C 16},
161(2000); B. Jurco et al., Eur. J. Phys.  {\bf C 17}, 521(2000);
B. Jurco et al., Eur. J. Phys. {\bf C 21}, 383(2001); X. Calmet et
al., Eur. J. Phys. {\bf C 23}, 363(2002).

\bibitem{so10} P. Aschieri et al., Nucl. Phys. {\bf B651},
45(2003).

\bibitem{su5} N. Deshpande and X.-G. He, Phys. Lett. {\bf B 533},
116(2002).

\bibitem{su3} X.-G. He, Eur. J. Phys. {\bf C 28}, 557(2003).

\bibitem{Oh}
 T.~Ohl and J.~Reuter,
 Phys.\ Rev.\ D {\bf 70} (2004) 076007
 [arXiv:hep-ph/0406098].


\bibitem{laser} I. Ginzburg et al., Nucl. Instrum. Meth. Phys.
Res. {\bf 202}, 57(1983).



\end{references}
\end{document}